          [2001/04/25 1.1 (PWD)]
\documentclass[a4paper,twocolumn]{esapub} 
\usepackage{natbib}
\usepackage{psfig}

\title{ALMA: Galaxies and AGN}
\author{C.L. Carilli}
\affil{National Radio Astronomy Observatory, Socorro, NM, USA, 87801}

\begin{document}


\maketitle

\begin{abstract}

With the ability to see into optically obscured regions with more than
an order of magnitude better sensitivity and spatial resolution
relative to current (sub)mm telescopes, ALMA will provide a unique
look into the physics of galaxy formation and active galactic
nuclei. In this paper I summarize the ALMA potential for studying star
forming galaxies and active galactic nuclei from the nearby universe
to the epoch of formation of the first luminous objects.

\end{abstract}

\section{Introduction}

Studies of the cosmic 'background' radiation at optical through far
infrared wavelengths show two peaks of roughly equal power
(Franceschini 2001). This background is not a true (ie. diffuse)
background, like the CMB, but arises from the summed light from
galaxies throughout the universe.  The optical peak corresponds to
direct starlight, while the peak in the FIR corresponds to star light
that has been reprocessed by dust. While such 'background'
calculations compress a tremendous amount of information, the basic
fact that the FIR and optical peaks are of similar strength implies
that roughly half the star light in the universe is absorbed by dust
and re-emitted in the infrared.

Far-IR through radio telescopes have the ability to see through the
dust in galaxies, into the regions of most active star formation.  An
excellent example of the affect of dust on our view of 'galaxy
formation' in the nearby universe is the galaxy IC 342 at a distance
of 2 Mpc. Figure 1 shows an overlay of the optical, mm continuum, and
CO emission from IC 342 (Meier \& Turner 2004). The optical emission
is dominated by a young star cluster at the galaxy center. However,
the mm continuum (corresponding to thermal emission from warm dust)
and molecular gas emission both peak at the ends of the inner bar,
indicating the sites of most active star formation. These regions are
highly obscured by dust in the optical.

Moving to high redshift, perhaps the best example of dust-obscured
galaxy formation is the brightest submm galaxy in the Hubble Deep
Field -- HDF850.1 (Hughes et al. 1998; Downes et al.  1999). Figure 2
shows the overlay of the optical and mm images of this field. Accurate
radio interferometry has shown that there is no optical counterpart to
HDF 850.1 down to the limit of the HDF. Subsequent imaging in the
near-IR has found a faint, red source (K = 23.5; I-K $> 5.2$) at the
radio/mm position, likely corresponding to a $z>3$ star forming galaxy
(Dunlop et al. 2004). If so, the intrinsic IR luminosity is $\sim
7\times 10^{12}$ L$_\odot$, with a star formation rate of a few
thousand M$_\odot$ year$^{-1}$.  Perhaps most impressively, this
single source (as opposed to 10$^4$ optical galaxies) could
potentially dominate the cosmic star formation rate density at $z>2$
in the HDF (Hughes et al. 1998).

\begin{figure}[htb]
\psfig{file=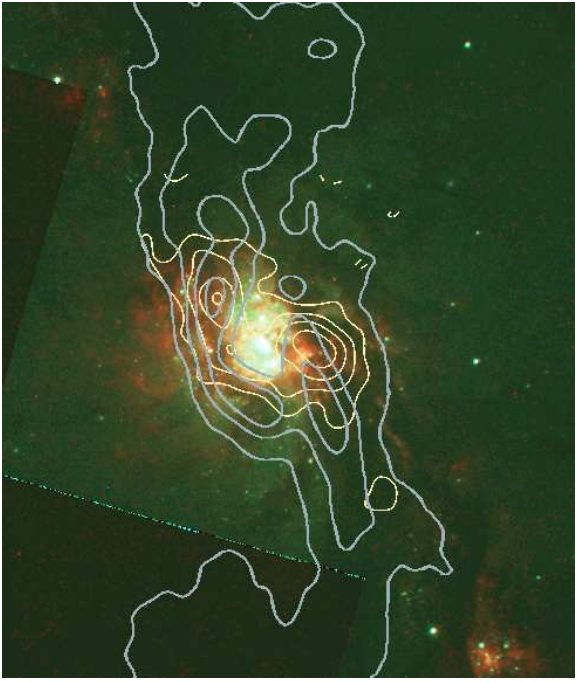,width=3in}
\caption{Images of IC342 in the optical (color), CO 1-0 (grey contours),
and mm continuum (white contours; from Meiers \& Turner 2004). The mm
continuum peaks indicate the 
regions of most active star formation, at the ends of the inner bar, 
are optically  obscured.
\label{IC342a}}
\end{figure}

\begin{figure}[htb]
\psfig{file=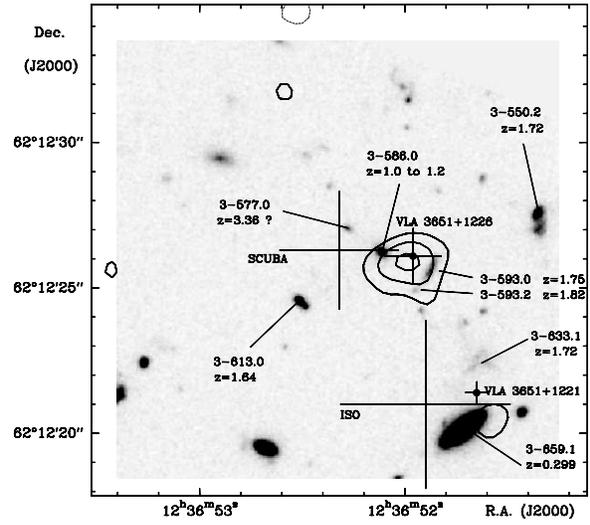,width=3in,angle=-90}
\caption{The contours show the PdBI image of the brightest 
(sub)mm source in the Hubble Deep Field, 
HDF850.1 (Downes et al. 1999), and the greyscale is the HST image. 
Accurate astrometry shows that the submm source has no optical counterpart
to the depth of the HDF.
\label{fig:HDF850.1}}
\end{figure}

A complementary viewpoint of the affect of dust on our understanding
of galaxy formation at high redshift is the study by Adelberger (2001)
of the UV through IR SEDs of high redshift galaxies.  He finds that
selection at rest frame uv wavelengths (ie. uv-dropouts or Ly-break
galaxies) is a very sensitive means of finding high redshift galaxies,
to well below L$_*$. However, he finds little correlation between uv
luminosity and bolometric luminosity. In other words, the higher
luminosity galaxies are also more heavily dust-obscured, such that,
while uv selection may find most of the star forming galaxies at high
redshift, the presence of dust complicates the physical analysis of
the cosmic star formation rate from uv selected samples.

\section{Enabling technologies}

Telescopes such as ALMA and Herschel are being designed to study 
dust obscured galaxy formation throughout the cosmos. What are the 
principle technological advances for ALMA that will lead to major
advances in the study of galaxy formation?

The most important advance for ALMA will be the two orders of
magnitude increase in sensitivity over existing mm arrays. Figure 3
shows the continuum spectrum of the active star forming galaxy Arp 220
in the radio through IR range, redshifted to z=2,5, and 8. Also shown
are the sensitivities of some current, or near-term, instruments, as
well as ALMA.  Current mm arrays such as the Plateau de Bure
interferometer and CARMA can, and have, detected ultra-luminous
infrared galaxies (ULIRGs, L$_{IR} > 10^{12}$ or star formation rates
$>$ a few hundred M$_\odot$ year$^{-1}$) to very high redshifts,
assisted by the large 'inverse-K' correction on the Rayleigh-Jeans
side of the dust spectrum. This large inverse-K correction can be seen
in the submm part of the spectrum in Fig. 3, where the flux density of
Arp 220 at a fixed observing wavelength is roughly constant for
redshifts from 0.5 to 8.  Such ULIRGs are rare in the nearby universe,
and are certainly not representative of what may be the normal star
forming galaxy population at high redshift, such as the Ly-break
galaxies, for which star formation rates are typically well below 100
M$_\odot$ year$^{-1}$ (Adelberger 2001). The radical increase in
sensitivity afforded by ALMA will enable study of even dwarf
starbursts ($\sim 10$ M$_\odot$ year$^{-1}$) out to extreme redshifts.

\begin{figure}
\psfig{file=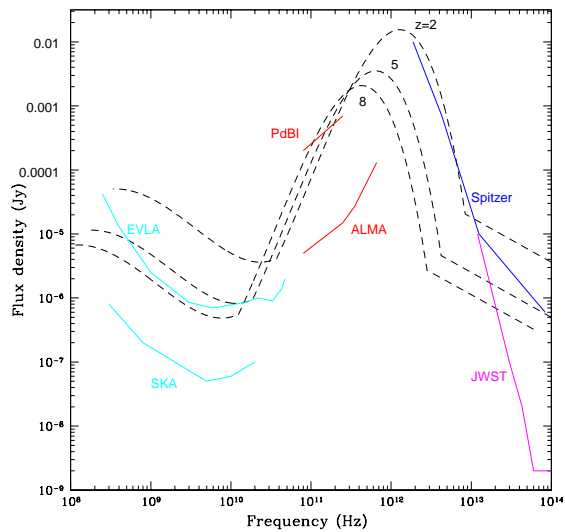,width=3in}
\caption{The dash lines show the spectrum of the active star forming
galaxy Arp 220 ($L_{FIR} = 1.3\times 10^{12}$ L$_\odot$) at
three  redshifts ($z=2$, 5, 8).
The solid lines show the rms sensitivity of current and
future instruments (in one transit) at cm through near-IR wavelengths.
\label{fig:arp220.sens.ps}}
\end{figure}

Figure 3 also shows the complementarity of ALMA with planned
facilities at other wavebands, such as the Square Kilomter Array and
the JWST.  Future instruments  will provide a pan-chromatic view of
star forming galaxies to extreme redshifts, into the epoch of 'first
light' in the universe (see section 5). Each of these wavebands
provides unique probes of the galaxy formation process, from the
non-thermal emission from star forming galaxies and AGN at cm
wavelengths, through the dust and molecular gas at mm and submm
wavelengths, to the stars, ionized gas, and AGN in the near-IR.

The second enabling technology for studying galaxy formation with ALMA
is the nearly two orders of magnitude improvement in spatial
resolution over existing connected-element mm arrays.  Figure 4 shows
the 'Walker diagram' of angular resolution vs. frequency for cm and mm
arrays.  ALMA will provide a resolution down to 10's of milliarcsecs
at 100's GHz, with brightness temperature sensitivity below 1 K. This
increase will reveal star formation on scales of Giant Molecular
Clouds (GMCs) out to 200 Mpc distances.

\begin{figure*}[htb]
\psfig{file=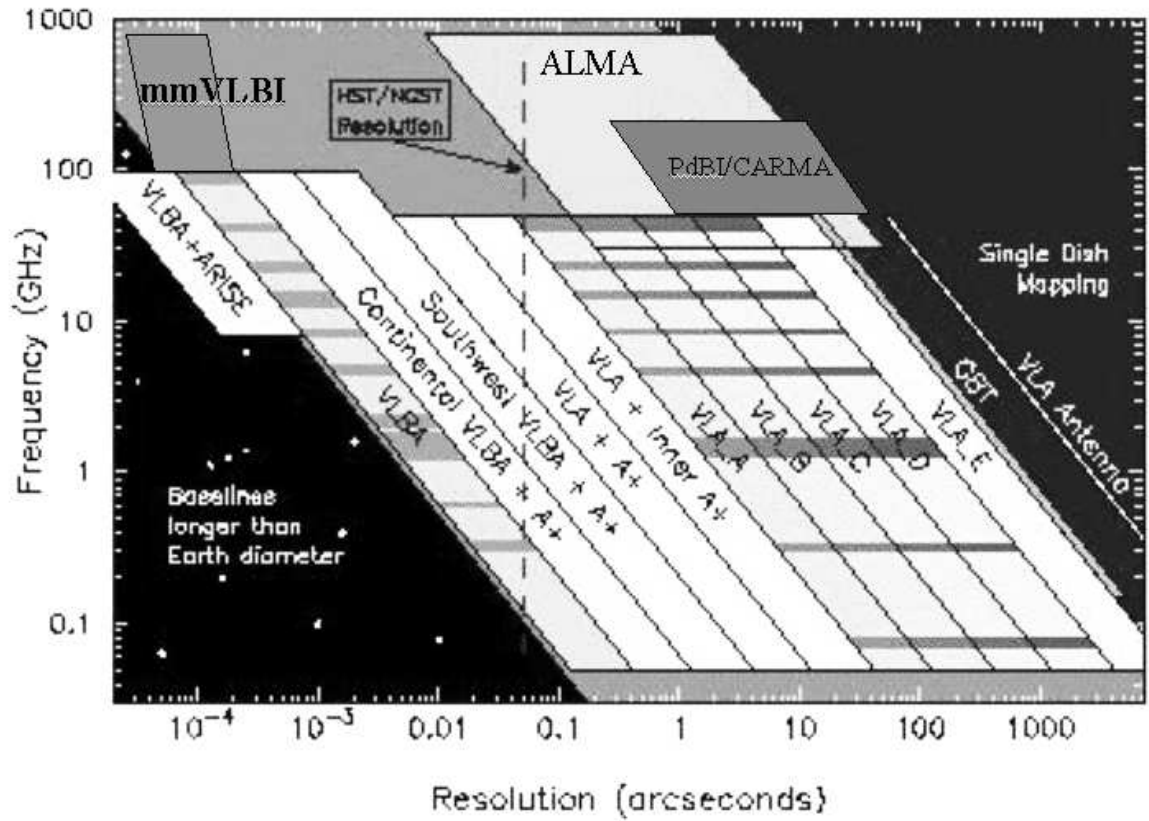,width=6.5in}
\caption{The 'Walker' diagram of resolution vs. frequency
for current and future radio and mm telescopes. 
\label{fig:Walker.ps}}
\end{figure*}

\section{Some current examples and where ALMA will take us}

\subsection{Nearby galaxies}

Consider the recent 'state-of-the-art' multi-transition study of
molecular gas in the very nearby star forming galaxy IC342 by Meier
and Turner (2004). Through detailed studies of both low and high
density gas tracers, as well as other physical diagnostics such as
photodissociation region tracers, they were able to delinate the
complex star formation processes throughout the disk, bar, and nucleus
of IC342 down to GMC scales. These processes include dense gas
associated with the most active star forming regions at the ends of
the inner bar, as traced by eg. HC$_3$N emission, and PDR regions
associated with the central star cluster, as traced by C$_2$H.

Similarly telling probes of gas dynamics on GMC scales in galaxies
have been obtained by Schinnerer et al. (2004) in NGC 6946 at 5.5 Mpc
distance.  These PdBI observations of CO 2-1 emission at 0.5$''$
resolution show clear signatures of changing gas dynamics, and likely
nuclear gas 'feeding', in the inner 10's of parsecs.

The important point is that these studies require physical resolutions
on scales of GMCs, and currently we are limited to galaxies not far
beyind the local group ($<$ few Mpc). ALMA will provide the resolution
and sensitivity to extend these studies out to 200 Mpc distance,
encompassing rich clusters such as Virgo and Coma, as well as extreme
starburst galaxies, such as Arp 220 and MRK 273, and luminous AGN,
such as Cygnus A, M87, and MRK 231.

\subsection{High redshift galaxies}

Studies of the evolution with redshift of the cosmic star formation
rate density (eg.  Blain et al. 2002) show a peak in the range $z=1.5$
to 3. One of the unanswered questions in this regard is the effect of
dust on this critical inventory of galaxy formation. Unfortunately,
current submm observations are limited to only the most extreme
systems at these distances (Fig 3). ALMA will push down to flux
densities of 10's of $\mu$Jy, ie. to normal star forming galaxies at
high redshift (star formation rates $\sim 10$'s M$_\odot$
year$^{-1}$), with sufficient resolution to avoid confusion limits
that plague single dish observations (Figure 5).  At this level the
(sub)mm source counts are comparable to the optical galaxy density
observed in the HDF (few$\times10^6$ deg$^2$). The key point is that
deep ALMA and optical surveys are clearly complementary, with optical
surveys dominated by galaxies at lower redshift ($z \le 1$), and ALMA
surveys revealing dusty star forming galaxies at higher redshift ($z >
1$).

\begin{figure}[htb]
\psfig{file=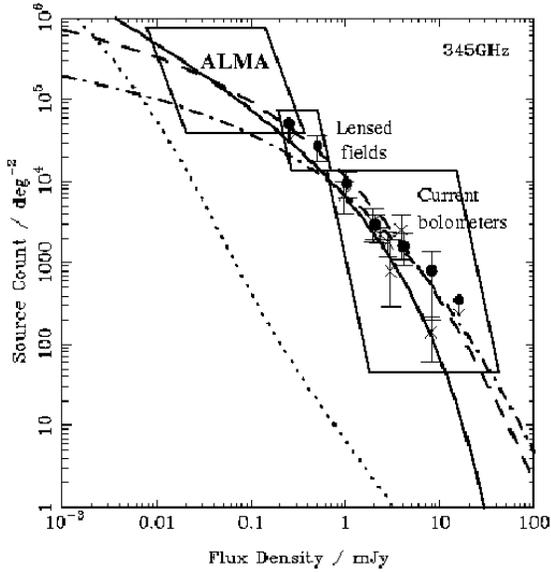,width=3in}
\caption{Source counts at 350 GHz (from Blain et al. 2002).
\label{fig:blain.cnts.ps}}
\end{figure}

In terms of molecular line studies, Cox (this volume) summarizes the
current situation for CO observations of low and high redshift
galaxies.  He shows the clear correlation between FIR luminosity and
CO luminosity for low redshift galaxies, consistent with a powerlaw of
index 1.7 (Gao \& Solomon 2004; Beelen et al. 2004; Carilli 2004). The
high redshift sources, which by necessity are also the highest
luminosity, are also shown, and interestingly, the high $z$ sources
continue the correlation to higher luminosity. Most of these sources
host known AGN, and yet they follow the same correlation of L$_{FIR}$
vs. L$'_{CO}$ as the low $z$ star forming galaxies. This correlation
could be used to argue that star formation is still the dominant dust
heating mechanism in the high $z$ sources (ie.  coeval starburst and
AGN).


\begin{figure*}[htb]
\psfig{file=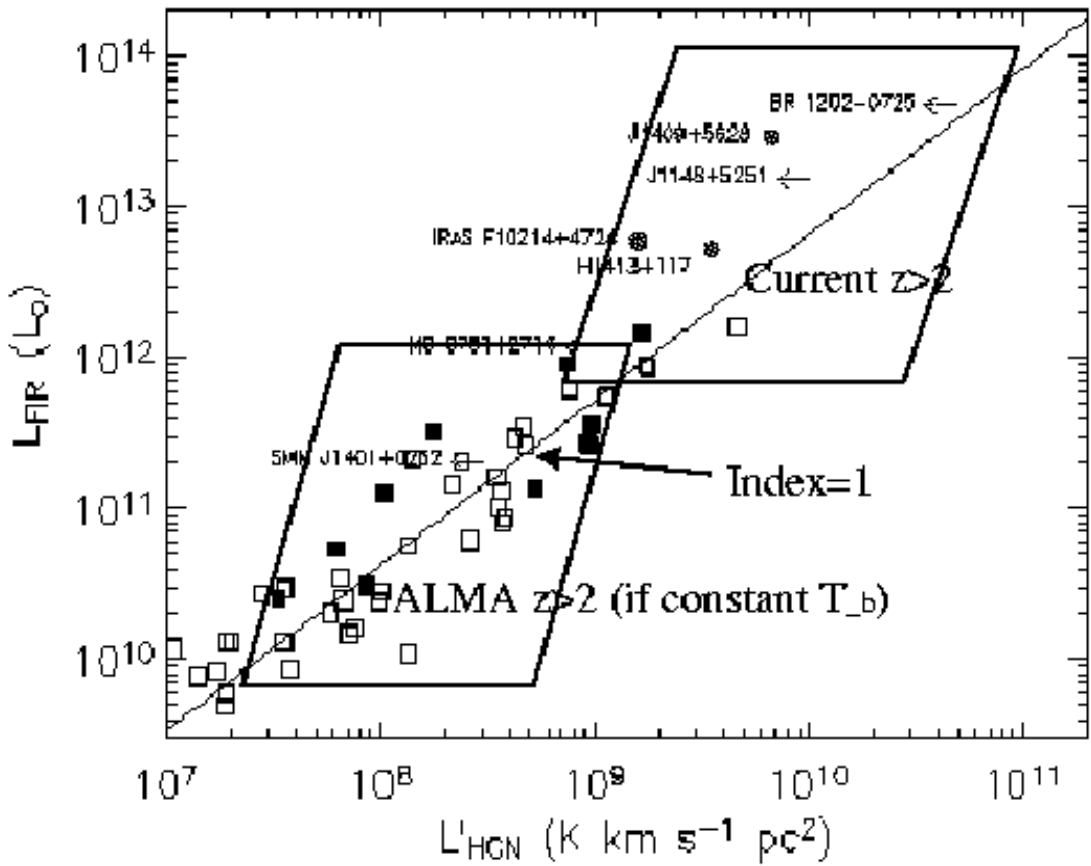,width=6in}
\caption{The correlation between
FIR and HCN luminosity for low $z$ (squares) and high $z$ (circles)
galaxies and AGN (Gao \& Solomon 2004;
Beelen et al. 2004; Carilli et al. 2004).
\label{fig:FIR-HCN.ps}}
\end{figure*}

The second strongest molecular line emission from star forming
galaxies is from the HCN molecule (Gao and Solomon 2004), with the HCN
emission typically being about 10$\%$ that of the CO luminosity,
although this fraction increases with increasing IR luminosity.
Figure 6 shows the L$_{FIR}$ vs. L$'_{HCN}$ correlation for nearby
galaxies, plus some recent measurements of high redshift galaxies
using the VLA (Carilli et al. 2004; Solomon et al. 2004).  HCN is in
important diagnostic, tracing dense gas directly associate with star
forming regions (critical density for excitation $\sim 10^5$
cm$^{-3}$), as opposed to CO which traces all the molecular gas
(critical density for excitation $\sim 10^3$
cm$^{-3}$). Interestingly, the HCN - FIR correlation is linear
(power-law index = 1), unlike the non-linear CO-FIR correlation. This
suggests that the FIR luminosity is linearly correlated with the dense
gas mass associated with active star forming clouds. The high $z$
sources generally fall along the linear correlation defined by the low
$z$ galaxies, again suggesting a similar dust heating mechanism
(ie. star formation). However, a number of the high $z$ sources have
only HCN lower limits, which would allow for some dust heating by the
AGN.

ALMA will push the studies of molecular line emission to the normal
galaxy population at high redshift, probing to Milky way type
molecular gas masses out to $z \sim 3$. Moreover, ALMA will provide
sub-arcsec imaging of the gas, to probe dynamics and dark matter on
kpc-scales.  Modelling by Blain (2001) has shown that blind surveys by
ALMA should detect 10's of galaxies per hour via their CO emission in
the redshift range 0.5 to 2.5. 

However, it should be noted that for dense gas tracers like HCN, ALMA
will be forced to study the higher order transitions at high redshift
(eg. a 90 GHz observing frequency corresponds to HCN 5-4 at $z =
4$). It is possible (likely?) that these transitions are sub-thermally
excited due to the very high critical densities involved. In this
case, study of the dense gas tracers at high redshift may be better
done using large area cm telescopes working in the 20 to 50 GHz range,
such as the EVLA, and eventually the SKA (Carilli \& Blain 2003).

One area where ALMA will clearly make fundamental breakthroughs is in
the study of the ISM submm cooling lines, such as C+ and CI (van der
Werf 1999, Papadopoulos et al. 2004). This area has been dissapointing
at high redshift, due to the relative weakness of the
strength of the C+ line in galaxies with warm IR spectra (Malhotra
this volume). This effect may be due to a decrease in the efficiency
of photoelectric heating by charged grains in regions of high
radiation fields (Wolfe 2004). By pushing down to normal galaxies,
where the C+ line is expected to dominate ISM cooling, ALMA will open
an exciting window into ISM physics in early galaxies.

\section{Complementarity}

ALMA will not work in a vacuum (unfortunately!), and it is important
to recognize  contributions from other telescopes. Indeed,
this conference is meant to highlight the dual roles of ALMA and
Hershel in the study of extragalactic astronomy, as can be seen in
these proceedings.  But in the mm regime itself, there will also be
large single dish telescopes providing complentarity to ALMA as well,
such as the LMT, GBT, APEX, ASTE...

One area where the single dish telescopes will contribute is through
very wideband spectroscopy (up to 32 GHz). Such wide band spectra
will have multiple transitions of CO, C+, HCN, and other molecules
in a single spectrum of a high z source, and hence provide
redshifts without having to rely on optical spectroscopy. 

A second area where single dish telescopes will contribute is with
large format bolometer cameras doing very wide field surveys to
sub-mJy sensitivity. The important point is that the small FoV of ALMA
makes very wide field surveys difficult. Indeed, future bolometer
cameras will be competitive with, or superior to, ALMA, in terms of
survey sensitivity for fields larger that $15'\times 15'$. Hence, one
can invision very wide field surveys with future sub-mm bolometer
cameras, as well as with radio and far-IR telescopes, to define
samples of interesting sources which can be followed-up with
sensitive, high resolution observations with ALMA to study the
detailed physics of the sources. Of course, for ultra-deep ($\mu$Jy),
narrow field studies of the submm source population, ALMA will be
incomparable.

\section{ALMA studies of cosmic reionization}

The discovery of the Gunn--Peterson absorption trough in the spectra
of the most distant quasars ($z>6$), corresponding to Ly $\alpha$
absorption by the neutral IGM, implies that we have finally probed
into the epoch of cosmic reionization (EoR; White et al. 2004).  The
EoR sets a fundamental benchmark in cosmic structure formation,
corresponding to the formation of the first luminous objects (star
forming galaxies and/or accreting massive black holes). Unfortunately,
G-P absorption during the EoR precludes observations of objects at
wavelengths longer than 0.9 micron. Hence study of the first galaxies
and AGN is the exclusive realm of near-IR to radio astronomy. The last
few years has seen a revolution in the number of objects discovered at
$z > 6$ using near-IR imaging and spectroscopy, including
star forming galaxies (Malhotra \& Rhoads 2004; 
Stanway et al. 2004; Hu \& Cowie 2002; Kodaira
et al.  2003) and AGN (Fan et al. 2003).

The recent discovery of molecular line emission, thermal emission from
warm dust, and radio syncrotron emission, from the most distant QSO
1148+5251 at $z=6.4$ (Bertoldi et al. 2003a,b; Walter et al. 2003;
Carill et al.  2004), implies very early enrichment of heavy elements
and dust in galaxies, presumably via star formation, within 0.8 Gyr of
the big bang (Figure 7).  The presence of a massive starburst in the
host galaxy of 1148+5251 is supported by the observed radio-FIR SED
(Figure 8), which follows the radio-FIR correlation for star forming
galaxies, with an implied star formation rate of order 10$^3$
M$_\odot$ year$^{-1}$ (Beelen et al. 2004).  Likewise, high resolution
imaging of the CO emission shows that the gas extends over
$\approx 1"$, with two peaks separated by $0.3"$, suggesting a merging
galaxy system (Figure 9). HST imaging shows that the optical QSO is
associate with the southern CO peak (White et al.  2004). And from the
gas dynamics, Walter et al. (2004) conclude that the supermassive
black hole forms prior to the formation of the stellar bulge in the
earliest AGN host galaxies.

\begin{figure}
\psfig{file=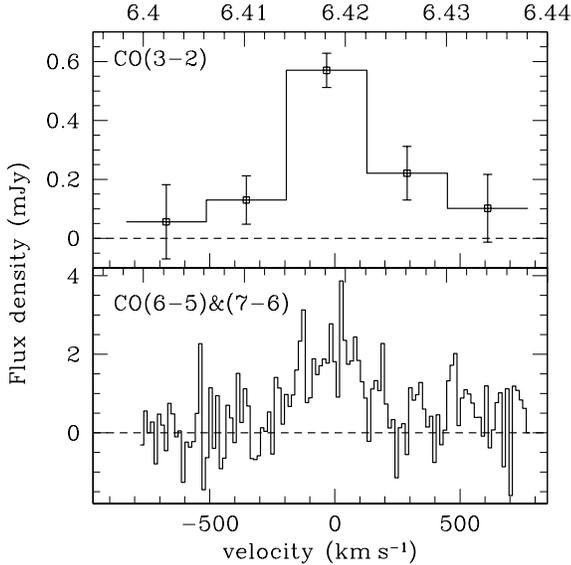,width=3in}
\caption{The CO line emission from the most distant QSO, 1148+5251 at
$z=6.42$ (Walter et al. 2003; Bertoldi et al 2003). The 3-2 line was
observed with the Very Large Array at 47 GHz, while the higher order
transitions were observed with the Plateau de Bure interferometer. The
implied molecular gas mass is $2.2\times 10^{10}$ M$_\odot$.
\label{fig:fig2_rev.ps}}
\end{figure}

\begin{figure}
\psfig{file=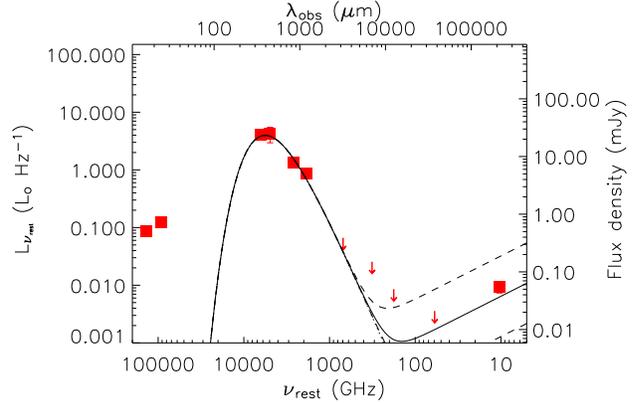,width=3.3in}
\caption{The radio through IR SED for the highest
redshift QSO, 1148+5251 at $z = 6.4$ (Beelen et al. 2004).
The curve shows the expected SED for a star forming galaxy.
\label{fig:Chris_1148_8GHz.eps}}
\end{figure}

\begin{figure}
\psfig{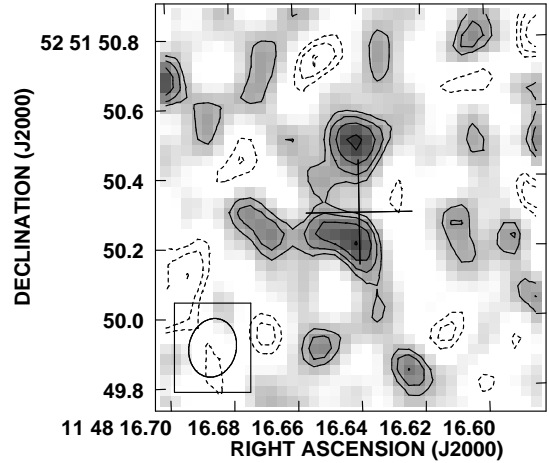}
\caption{A high resolution ($0.15"$) VLA image of the CO 3-2 emission
from the highest redshift QSO, 1148+5251 at $z = 6.4$ (Walter et al. 2004).
HST imaging of the optical QSO shows that it is associated with
the southern CO component (White et al. 2004). 
\label{fig:f3.eps}}
\end{figure}

These studies of 1148+5251 demonstrate the power of mm line and
continuum studies of the earliest galaxies and AGN.  Unfortunately,
the observations of 1148+5251 stretch current instrumentation to the
extreme limit, such that only rare and pathologic objects are
detectable, ie. hyperluminous IR galaxies with $L_{FIR} > 10^{13}$
L$_\odot$.  The two orders of magnitude increase in sensitivity
afforded by ALMA will enable study of the molecular gas and dust in
the first 'normal' galaxies within the EoR.  Such studies will reveal
the physics and chemistry of molecular gas reservoirs required for
star formation, and provide a unique probe of gas dynamics and
dynamical masses of the first galaxies. In parallel, radio continuum
studies with nJy sensitivity in the frequency range 1 to 10 GHz with
the EVLA and, eventually, the SKA, will present a dust-unbiased view
of star formation in these systems.

As a concrete example of the types of objects that might be studied,
consider the galaxies being discovered in Ly$\alpha$ surveys at $z
\sim 6$. The typical UV luminosity is a few$\times10^{10}$ L$_\odot$.
Making the standard factor five dust correction for typical high
redshift star forming galaxies (ie. Ly-break galaxies) implies an FIR
luminosity $\sim 10^{11}$ L$_\odot$.  The predicted thermal emission
from warm dust at 250 GHz is 25 $\mu$Jy, which can be detected at
4$\sigma$ with ALMA in one transit (6hrs). We expect one or two of
these objects in every ALMA FoV.

Lastly, an important aspect of the molecular line observations of
galaxies within the EoR is that they give the most accurate redshifts
(by far) for the host galaxies. Typical high ionization broad metal
emission lines from QSOs are notoriously uncertain in terms of the
host galaxy redshifts, with offsets typically on the order of 10$^3$
km s$^{-1}$ (Richards et al. 2002), while Ly $\alpha$ emission lines
are affected severely by absorption.  Accurate host galaxy redshifts
are crucial in the calculation of the size of cosmic Stromgren spheres
around objects within the EoR, since these sizes are derived from the
redshift difference between the host galaxy and the on-set of GP
absorption (Wyithe \& Loeb 2004). The sizes of these ionized regions
have been used to constrain the IGM neutral fraction (Wyithe et al.
2005), setting a lower limit to the neutral fraction of 0.1 at $z \sim
6.4$, two orders of magnitude more stringent than the lower limit set
by the GP effect.

\section{Millimeter VLBI Observations of the Galactic Center}

A final program we consider is (sub)mm VLBI observations of the
supermassive black hole at the Galactic center, including (phase
array) ALMA as the most sensitive element in the array. Other possible
elements include the LMT, CARMA, JCMT (or CSO or SMA), the HHT, PdBI,
and the IRAM 30m. These observations will allow for imaging at $\sim
10\mu$as resolution, well matched to the scale of the expected general
relativistic shadow of the SMBH in Sgr A* (Falcke et al. 2000).

\begin{figure*}[htb]
\psfig{file=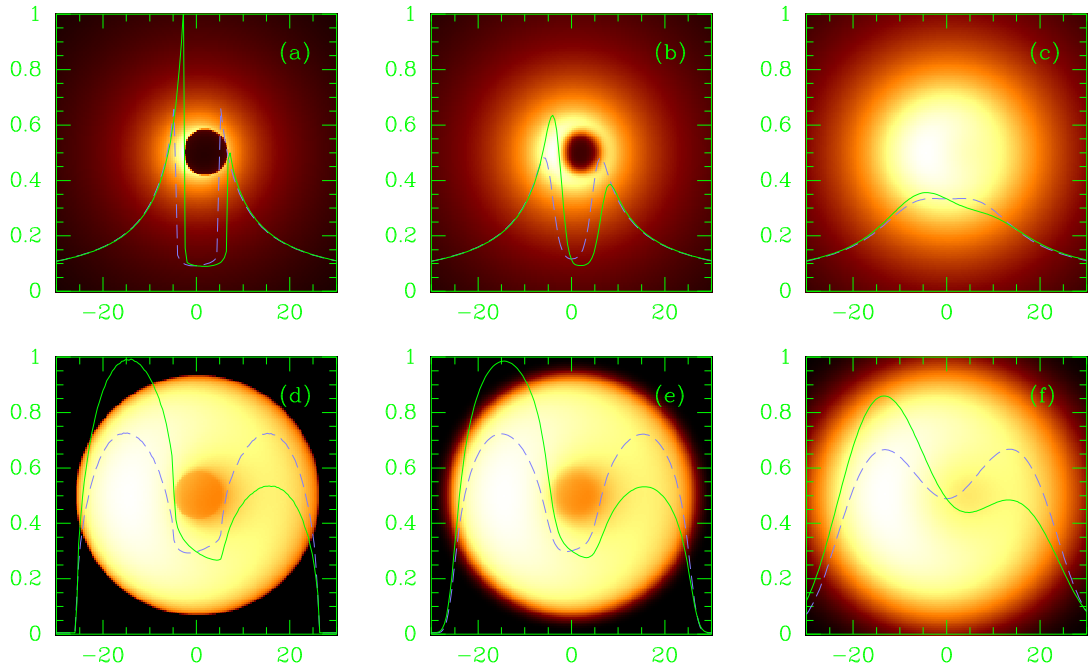,width=6in}
\vskip -1.5in
\caption{Simulations of the expected general relativistic shadow
of the Galactic center supermassive black hole as seen in 
(sub)mm VLBI images at 10's $\mu$as resolution (Falcke et al. 2000).
The predicted signature depends strongly on whether the hole
is rotating (upper frames) or not (lower frame), ie. Kerr or 
Schwarzschild, since rotation affects
the radius of the last stable orbit.  The left frames are
the model. The center frames are for observations at 0.6mm,
including scattering, and the right frames are at 1.3mm. The tick
marks along the X axis are in $\mu$as. 
\label{fig:sgra.ps}}
\end{figure*} 

Figure 10 shows the expected signature of the black hole on the
non-thermal brightness distribution at (sub)mm wavelengths.  These
observations will provide the ultimate evidence for the existence of a
SMBH at the Galactic center, provide a fundamental test of strong
field GR, and are the most direct method for separating a Kerr
(ie. spinning) from a Schwarzschild black hole.  At a minimum, the
sensitivity per baseline is adequate to perform model fitting on
relatively short timescales (minutes), while the VLBI array itself has
enough antennas to provide both closure amplitude and phases, and
hence should be adequate for hybrid imaging of the GR shadow of Sgr
A*. The existence of reasonable mm-VLBI calibrators (eg NRAO 530) in
the vicinity of Sgr A* will allow for phase-referenced fringe fitting,
although the source itself is strong enough, and the UV coverage dense
enough, to allow for hybrid mapping as well.  

The source Sgr A* has been detected at 220 GHz on the PdBI -- Pico
Veleta baseline (resolution = 300 uas) with a flux density of 2.0 Jy,
and an upper limit to the size of order 100 uas (Krichbaum et
al. 1998). The proposed observations will have more than an order of
magnitude better resolution, more than two orders of magnitude better
sensitivity, and, again, enough antennas to perform proper imaging of
the general relativistic shadow of Sgr A*.

\section*{Acknowledgments}

The National Radio Astronomy Observatory is operated by Associated
Universities Inc., under cooperative agreement with the National
Science Foundation. I would like to thank my collaborators (F.Betoldi,
F. Walter, P. Cox, A. Beelen, P. Solomon, P. van den Bout, etc...) for
allowing me to reproduce some of our recent work, and H. Falcke,
C. Walker, D. Meier, J. Turner, D. Downes, A. Blain for use of other
figures.


\begin{thebibliography}{}

\bibitem{}Adelberger, K. 2001, in {\sl Starburst galaxies near and far},
eds. Tacconi \& Lutz, (Springer: Heidelberg), p. 318

\bibitem{}Blain, A. et al. 2002, Phys. Rep., 369, 1

\bibitem{}Blain, A. 2001, in {sl Science with the Atacama Large Millimeter Array},
ed. A. Wootten, (ASP: San Francisco), p. 261

\bibitem{}Beelen, A. 2004, PhD Thesis, U.Paris-Sud

\bibitem{}Bertoldi, F., Cox, P., Neri, R. et al. 2003,
A\&A, 409, L47

\bibitem{}Bertoldi, F., Carilli, C., Cox, P. et al. 2003,
A\&A, 406, L15

\bibitem{}Carilli, C.L. et al. 2004, ApJ, in press

\bibitem{}Carilli, C. \& Blain, A. 2002, ApJ, 569, 605

\bibitem{}Carilli, C., Bertoldi, F., Walter, F. et al. 2004b, in Multiwavelength
AGN Surveys, eds. Maiolino and Mujica (World Scientific), in press
(astroph/0402573)

\bibitem{}Downes, D. et al. 1999, A\& A, 347, 809

\bibitem{}Dunlop, J. et al. 2004, MNRAS, 350, 769

\bibitem{}Fan, X., Strauss, M., Schneider, D. et al.2003, AJ,  125, 1649

\bibitem{}Falcke, H., Melia, F., Agol, E. 2000, 528, L13

\bibitem{}Franceschini, A. 2001, IAU Symp. 204, ed. Harwit, p. 283

\bibitem{}Gao, Y. \& Solomon, P. 2004, ApJS, 152, 63

\bibitem{}Gao, Y. \& Solomon, P. 2004, ApJ, 606, 271

\bibitem{}Hu, E., Cowie, L., McMahon, R. et al. 2002, ApJ, 568, L75

\bibitem{}Hughes, D. et al. 1998, Nature, 394, 241

\bibitem{}Kodaira, K., Taniguchi, Y.Kodaira, K., Taniguchi, Y., Kashikawa, N. 2003,
  PASJ, 55, L17

\bibitem{}Krichbaum et al. 1998 335, L106

\bibitem{}Malhotra, S. \& Rhoads, J. 2004, ApJ, in press

\bibitem{}Meier, D. \& Turner, J. 2004, ApJ, in press

\bibitem{}Papadopoulos, P. \& Greve, T. 2004, ApJ, 615, L29

\bibitem{}Richards, G.T., vanden Berk, D., Reichard, T. 2002,   AJ, 124, 1

\bibitem{}Schinnerer E., et al. 2004, A\& A, in prep

\bibitem{}Stanway, E., Glazebrook, K., Bunker, A. et  al. 2004, ApJ 604, L13

\bibitem{}van der Werf, P. 1999, in {\sl Highly redshifted radio lines}, eds.
Carilli et al., (ASP: San Francisco), p. 91

\bibitem{}Walter, F., Bertoldi, F., Carilli, C. et al. 2003,  Nature, 424, 406

\bibitem{}White, R., Becker, R., Fan, X., Strauss, M. 2003,  AJ, 126, 1

\bibitem{}Withe, A. \& Loeb, L. 2004, Nature, 427, 815

\bibitem{}Wolfe, A., Prochaska, J., Gawiser, E. 2003, ApJ, 593, 215

\end{thebibliography}
\end{document}